\documentclass[prb,twocolumn,superscriptaddress,floats]{revtex4}

\usepackage[dvips]{graphicx}
\usepackage{amsmath}
\usepackage{booktabs}
\usepackage{dcolumn}
\usepackage{sidecap}

\usepackage{rotating}

\newcolumntype{d}{D{.}{.}{-1}}

\newcommand{\re}{$RE$TiO$_3$}
\newcommand{\tit}{TiO$_3$}
\newcommand{\grad}{\ensuremath{^\circ}}

\newcommand{\TN }{$T_\text{N}$}
\newcommand{\TC }{$T_\text{C}$}
\newcommand{\eg}{e$_{g}$}
\newcommand{\tg}{t$_{2g}$}

\newcommand{\rei}{rare-earth ionic radius}

\newcommand{\tioct}{TiO$_6$}
\newcommand{\tioc}{TiO$_6$}
\newcommand{\roo }{$\Delta(O2-O2)$}
\newcommand{\rooD }{$\Delta(O2-O2)$}
\newcommand{\rmoD }{$\Delta(M-O2)$}

\newcommand{\TNSm} {45 K}

\begin{document}


\title{Magnetoelastic coupling in \re\ ($RE$ = La, Nd, Sm, Gd, Y)}

\author{A. C. Komarek }
\affiliation{II. Physikalisches Institut, Universit\"{a}t zu
K\"{o}ln, Z\"{u}lpicher Str. 77, D-50937 K\"{o}ln, Germany}
\author{H. Roth}
\author{M. Cwik }
\author{W.-D. Stein }
\author{J. Baier}
\author{M. Kriener }
\affiliation{II. Physikalisches Institut, Universit\"{a}t zu
K\"{o}ln, Z\"{u}lpicher Str. 77, D-50937 K\"{o}ln, Germany}
\author{F. Bour\'ee }
\affiliation{Laboratoire L\'eon Brillouin, CEA-CNRS, CE-Saclay, F-91191
Gif-sur-Yvette, France }
\author{T. Lorenz }
\affiliation{II. Physikalisches Institut, Universit\"{a}t zu
K\"{o}ln, Z\"{u}lpicher Str. 77, D-50937 K\"{o}ln, Germany}
\author{M. Braden}
\email{braden@ph2.uni-koeln.de}
\affiliation{II. Physikalisches
Institut, Universit\"{a}t zu K\"{o}ln, Z\"{u}lpicher Str. 77,
D-50937 K\"{o}ln, Germany}

\date{\today}

\pacs{PACS numbers:}


\begin{abstract}

A detailed analysis of the crystal structure in $RE$TiO$_3$ with
$RE$ = La, Nd, Sm, Gd, and Y reveals an intrinsic coupling between
orbital degrees of freedom and the lattice which cannot be fully
attributed to the structural deformation arising from bond-length
mismatch. The TiO$_6$ octahedra in this series are all irregular
with the shape of the distortion depending on the $RE$ ionic
radius. These octahedron distortions vary more strongly with
temperature than the tilt and rotation angles. Around the Ti
magnetic ordering all compounds exhibit strong anomalies in the
thermal-expansion coefficients, these anomalies exhibit opposite
signs for the antiferromagnetic and ferromagnetic compounds.
Furthermore the strongest effects are observed in the materials
close to the magnetic cross-over from antiferromagnetic to
ferromagnetic order.

\end{abstract}
\maketitle

\section{Introduction}

The rare-earth titanates (\re) with a distorted perovskite
structure have a single electron in the \tg \ orbitals of the
Ti-$3d$ shell and are all Mott insulators \cite{imada}. Therefore,
they can be considered as the one-electron counterpart to the
cuprates with a single hole in the $3d$ shell. Upon decreasing the
ionic radius in the rare-earth $RE$ series from La to Y
\cite{shannon}, the calculated one-electron bandwidth gets smaller
\cite{okimoto} and the deviation of the Ti-O-Ti bond angle from
180\grad\ increases \cite{maclean}. Concomitantly, the Ti magnetic
order changes from G-type antiferromagnetism with the magnetic
moment in \emph{a} direction ($RE$ = La,...,Sm) to ferromagnetism
with the magnetic moment in \emph{c} direction ($RE$=Gd,...,Yb and
Y)
\cite{cwik,amow1,onoda9a,zhou,onoda9b,keimer,meijer,hays,goral,amow2,akimitsu,garrett}
. These apparently distinct magnetic ordering schemes, however,
belong to the same irreducible representation, as it has been
clarified long ago \cite{bertaut} for the structure type of
GdFeO$_3$ to which all $RE$\tit \ belong. Due to the low symmetry
of the crystal structure, the G-type antiferromagnetic component
in \emph{a} direction is always coupled with an A-type
antiferromagnetic moment in \emph{b} direction and a ferromagnetic
component in \emph{c} direction. Thus the change from
antiferromagnetic to ferromagnetic order of the Ti moments in the
series of \re\ is just a redistribution of the ordered moment
between the three different components ($G_x$, $A_y$ and $F_z$)
within the same magnetic symmetry.

Tremendous efforts have been made to understand the magnetic
properties of these titanates during the last years
\cite{solovyev1,solovyev2,khaliullin,khaliullin2,pavarini,moshizuki,
schmitz1,schmitz2}. An orbital liquid model has been proposed in
the idea that the splitting between the \tg \ states would be
sufficiently low to allow orbital fluctuations to play a dominant
role in the physical properties \cite{khaliullin,khaliullin2}. On
the other hand, in a more conventional scenario, it is argued that
orbital fluctuations would be efficiently suppressed by the \tg
-level splitting of the order of 200\ meV
\cite{pavarini,cwik,moshizuki,haverkort,rueckkamp}. Taking account
of the anisotropic orbital arrangement Schmitz and
M\"uller-Hartmann could explain nearly all the details of the
magnetic order in La\tit \ including the measured dispersion of
the magnetic excitations\cite{schmitz1,schmitz2}.

In addition to the Ti magnetism, magnetic order of the $RE$
moments occurs at lower temperature. For the lighter $RE$ ions
($RE$ = Ce, ... , Sm) the $RE$ moments order antiferromagnetically
below $T_{N2}<T_{N1}$ (with $T_{N1}$ being the N\'{e}el
temperature for $G_x$-type antiferromagnetic ordering at the Ti
sites). For the \re\ compounds studied in this work belonging to
this first group ($RE$ = Nd, Sm) an antiferromagnetic ordering of
$F_y/C_z$-type is observed at the $RE$ sites \cite{amow1, amow2}.
For the \re\ with heavier $RE$ ions ($RE$ = Gd,...,Yb) there are
two basic magnetic structures. In the first class, the $RE$
magnetic moments order ferromagnetically and couple antiparallel
to the ferromagnetic component of the Ti sites ($F_z$). In the
second class ($RE$=Tb, Dy and Ho) the magnetic moments at the $RE$
sites have both ferromagnetic and antiferromagnetic components in
the $a,b$ plane \cite{turnerJSSC,zhou,turnerJMMM, greedan,
heinrich}. Among the heavier $RE$\tit \ materials only Gd\tit \
has been studied in this work, it exhibits ferromagnetic Gd-moment
ordering (first group). These complex $RE$ magnetic structures,
however, seem to yield little insight to the Ti magnetism and the
associated interaction energies are smaller than those in-between
the Ti's except in the compounds where the Ti-Ti interaction is
strongly suppressed.

\begin{figure}[t]
\begin{center}
\includegraphics*[width=0.6\columnwidth]{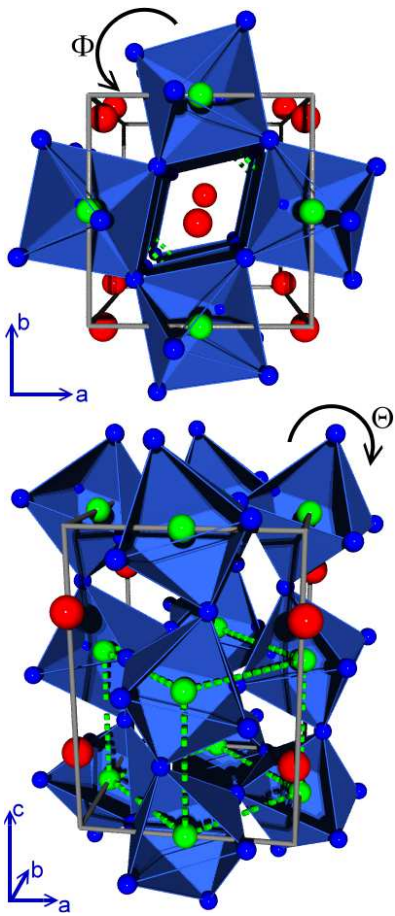}
\end{center}
\caption{ (color online) Crystal structure of $RE$TiO$_3$ with
corner sharing \tioct\ octahedra in \emph{Pbnm} notation. Dashed
lines show the cubic unit cell of the undistorted structure and
solid lines show the orthorhombic unit cell. \emph{Green:}
Ti-ions, \emph{blue:} O-ions and \emph{red:} $RE$-ions.}
\label{fig1}
\end{figure}
\par

The GdFeO$_3$-type structure of the rare-earth titanates differs
from the ideal, cubic perovskite structure (with space group
\emph{Pm$\overline{3}$m}) by a rotation of the octahedra around
the $c$ axis and by an additional tilt of the \tioct \ octahedra
around the cubic $[110]$ axis. While the octahedral tilts are
opposite for all neighboring \tioct \ octahedra, the octahedra are
rotated in the same sense along the \emph{c} direction and
opposite only in the \emph{ab} planes.  The combination of tilt
and rotation distortion reduces the symmetry to the orthorhombic
space group \emph{Pbnm} and enlarges the unit cell to a
$\sqrt{2}\times \sqrt{2}\times 2$   cell rotated by 45\grad\
around the \emph{c} axis (see Fig \ref{fig1}). In this space group
internal distortions of the \tioct \ octahedra can occur without a
further symmetry breaking. The most famous example for such
additional distortions is LaMnO$_3$ where the antiferroorbital
order of the \eg  \ electrons coexists with the tilt and the
rotation\cite{mn1,mn2,mn3}. Qualitatively similar octahedron
distortions have been observed for all rare earth titanates
studied in this work \cite{maclean,cwik}. For an undistorted
octahedron, one expects a negative orthorhombic splitting
$\varepsilon=(a-b)/(a+b)$ since the octahedral tilt takes place
around the orthorhombic \emph{b} axis (\emph{Pbnm}). But due to an
elongation of the \tioct \ octahedraon basal plane along the
orthorhombic \emph{a} direction the orthorhombic splitting in
La\tit\ becomes even positive. In the series of \re\ the type of
octahedral distortion changes its character with decreasing \rei .
For La\tit \ the O2-O2 distances (O1 denotes the apical oxygen
displaced from the Ti sites along $c$ and O2 the basal in-plane
oxygen) along the octahedron edges differ strongest resulting in a
rectangular shape of the basal oxygen planes. In contrast, in
Y\tit \ there is little difference in these O2-O2 distances but
the two in-plane Ti-O2 distances split. These distortions together
with the anisotropic Ti-$RE$ coordination lead to a splitting of
the \tg \ energy levels with a different character of the lowest
state. Interestingly the orbital arrangement resulting from these
two distortions is totally different. Within a single layer
parallel to the $a,b$ plane, the variation of the edge lengths
yields a ferroorbital configuration whereas the splitting of the
Ti-O2 distances yields an antiferroorbital configuration
\cite{cwik}. According to Goodenough-Kanamori rules, these
distinct orbital ordering schemes agree with the antiferromagnetic
and ferromagnetic Ti ordering in La\tit \ and in Y\tit ,
respectively.

Although there is no doubt that the octahedron distortions in the
$RE$\tit \ series are coupled with the magnetic ground state, it
is less obvious whether these effects should be considered as
resulting from orbital ordering. There is no symmetry breaking in
the titanate case as the orbital distortions are embedded in the
crystal structure deformation arising from tilt and rotation.
Pavarini et al. \cite{pavarini} argue that the values of the
tilting and the concomitant shift of the $RE$ sites would
determine the splitting of the orbital levels \cite{pavarini}.
However, the lattice parameters and the octahedron deformation in
La\tit \ exhibit clear anomalies around the magnetic ordering
suggesting that there is an intrinsic coupling between orbital
degrees of freedom and the crystal structure
\cite{cwik,hemberger}. It is the aim of the present work to
further elucidate this orbital-lattice coupling in the $RE$\tit \
series and to separate the intrinsically electronic effects from
tilt and rotation distortions. For this purpose we have analyzed
the known crystal structure of oxides showing the same GdFeO$_3$
structure type but having no electron in the $3d$ shell.
Furthermore we have studied the temperature dependence of the
crystal structure in $RE$\tit \ for $RE$ = La, Sm, Nd, Gd, and Y
in great detail. In all compounds we find evidence for a direct
orbital-lattice coupling in addition to that arising from the
distortions.

\section{Experimental}

Single crystals of \re\ were grown using a floating-zone image
furnace, as described elsewhere \cite{roth}. Parts of them were
crushed in order to obtain powder samples of high quality.
Magnetic ordering temperatures of the Ti ions were determined by a
SQUID or a vibrating sample magnetometer. The magnetic ordering of
the Ti depends sensitively on the stoichiometry
$RE_{1-x}$TiO$_{3+\delta}$. In order to obtain crystals with a
high transition temperature it is necessary to tune the parameters
during the growth process. In this work several crystals of the
same material were grown and analysed; the highest
T$_N$/T$_C$-values, we find, agree with or are higher than those
reported in a recent study\cite{zhou}. The N\'{e}el temperatures
of the antiferromagnetic compounds are 146 K for La\tit; 94 K and
81 K for two samples of Nd\tit\ and 54 K and 48 K for two crystals
of Sm\tit. In the case of Nd\tit\ the high T$_N$-sample was used
for diffraction and thermal expansion studies and the lower
T$_N$-sample only for diffraction studies. For Sm\tit, the 54 K
sample was studied in thermal expansion and the 48 K sample by
diffraction techniques. The Curie temperatures for the
ferromagnetic titanates are 36 K ($RE$=Gd) and 29 K and 25 K for
two Y\tit\ samples used for diffraction and thermal expansion
measurements, respectively. The Several diffraction techniques
were applied to analyze the temperature dependencies of the
structural parameters. A Siemens D5000 diffractometer equipped
with a one-dimensional detector was used for powder X-ray
diffraction with Cu-K$_\alpha$ or Cr-K$_\alpha$ radiation. These
as well as the neutron powder diffraction data were analyzed by
the Rietveld technique, but the laboratory X-ray powder
diffraction does not allow one to obtain the oxygen structural
parameters. In order to perform full structure analyzes powder
neutron-diffraction patterns were recorded on the high-resolution
3T.2 diffractometer at the Orph\'{e}e reactor in Saclay
($\lambda=1.2251$\AA). The full crystal structure is also
determined by single-crystal X-ray diffraction studies on a
\emph{Bruker X8 Apex} CCD diffractometer using Mo-K$_\alpha$
radiation. Absorbtion correction for spherical samples was
performed with the standard $Bruker$ software ($scale$) and
isotropic extinction correction was applied during the refinement
on F$^2$ with $Jana2000$\cite{jana2000}. Some of the crystals were
twinned. Due to very similar lattice parameter in La\tit, the
components of all twin domains could be integrated together and
refined with different volume fractions for all twin domains
within $Jana2000$. The corresponding twin laws result from a
threefold rotation about the [111] axis of the 8 \AA\ pseudocube
combined with a reflection across the (1-10) plane. In the other
$RE$\tit, no twinning was observable or the contributions of the
different twins could be separated during integration procedure
thus obtaining a dataset only for the major twin domain. This was
the case in Nd\tit, where the reflections of different twin
domains which could not be separated from each other were
dismissed (see also Tab. \ref{tableX}). These minor twin domains
have only small volume fractions of roughly 10\% of the total
sample volume. The structural results of our single crystal X-ray
and powder neutron-diffraction experiments are summarized in Tab.
I \& II. Thermal-expansion coefficients were measured for
single-crystalline Y\tit , Sm\tit , and Nd\tit \ on a
high-resolution capacitance dilatometer\cite{lorenz97a}. For Y\tit
\ we furthermore measured the magnetostriction along the $a$
direction in magnetic fields up to 14 T.

\section{Results and Discussion}

\subsection{Analysis of known GdFeO$_3$-type structures}

Since it is not obvious in the $RE$\tit  \ series to separate the
intrinsic orbital-ordering effects from those imposed by the
severe tilt and rotation distortions, it appeared interesting to
analyze the published crystal structures of metal oxides AMO$_3$
in  GdFeO$_3$-type structure which posses an empty $3d$ shell.
Using the Inorganic Crystal Structure Database (ICSD) \cite{icsd}
we have studied the compounds : CaTiO$_3$, CaZrO$_3$, EuScO$_3$,
GdGaO$_3$, GdScO$_3$, HoAlO$_3$, LaGaO$_3$, NdGaO$_3$, PrGaO$_3$,
SmAlO$_3$, SrZrO$_3$, TbAlO$_3$ with respect to their tilting
($\Theta$) and rotation ($\Phi$) distortions and to the
deformation of the metal-oxygen octahedron. \cite{TbAlO3} We
discuss the relative difference between the two M-O2 distances,
\rmoD, and that between the two O2-O2 basal-plane edge lengths,
\rooD.
\begin{figure}[t]
\begin{center}
\includegraphics*[width=0.85\columnwidth]{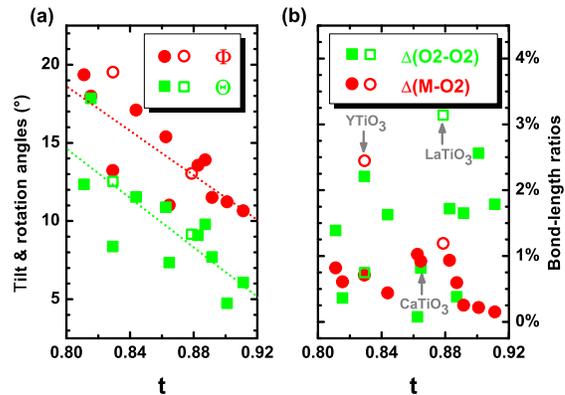}
\end{center}
\caption{(color online) (a) The octahedral tilt $\Theta$ and the
rotation $\Phi$ versus the tolerance factor t for various d$^0$
systems with perovskite structure (space group number 62, Pbnm
\cite{tables}). (b) Difference of both basal M-O2 bond lengths and
difference of both O2-O2 bond lengths versus t. (Open symbols
denote the values for La\tit\ and Y\tit; lines are linear fits to
the data.)} \label{figLit}
\end{figure}
The analysis of the published structures is resumed in Fig.
\ref{figLit}. The tilt and rotation deformations arise from the
bond-length mismatch between the M-O and A-O bonds which in the
case of an ideal perovskite would be forced into a ratio of
$\sqrt{2}$. With the ionic radii \cite{shannon} one calculates the
tolerance factor $t=(r_A + r_O)/[\sqrt{2}\cdot(r_M + r_O)]$ with
values much below $t=1$ indicating a strong mismatch due to too
small A ionic radii. Fig. \ref{figLit} $(a)$ shows the tilt and
rotation angles as a function of the tolerance factor yielding the
expected increase of both distortions towards smaller tolerance
factors. Also the ratio between the tilt and rotation angles seems
to be fully determined by the tolerance factor. One may note that
La\tit\ and Y\tit\ match these relations. In Fig. \ref{figLit}
$(b)$ we show the \rmoD\ and \rooD\ values again as a function of
the tolerance factor. For the \rmoD\ bond-distance ratio there is
no clear tolerance-factor dependence visible, but the deformation
of about 2.5\% seen in Y\tit\ is much larger than those in all the
others materials which vary only between 0\% and 1\%. This gives
the first evidence that the \rmoD \ deformation is not purely a
consequence of the tilt and the rotation in the titanates.
Concerning the edge-length deformation characterized by \rooD ,
the deformation of about 3\% seen in LaTiO$_3$ is strongest but
there seems to be a sizeable distortion also in the series without
$3d$ electrons. This deformation is intimately coupled with the
orthorhombic splitting. Tilting of a rigid octahedron around $b$
will always yield a smaller $a$ lattice parameter, whereas some of
the materials, including LaTiO$_3$, show the opposite. This
reversed orthorhombic splitting, however, appears only for small
tilt angles. It seems to result from the interactions between the
A site and the apical oxygens. One of these bond distances is
significantly shortened by a rigid tilt. The systems may therefore
elongate perpendicular to the tilt axis in order to reduce this
effect. This will yield a weak elongation of the octahedron basal
plane along $a$ and eventually the reversed orthorhombic
splitting. Although LaTiO$_3$ exhibits the strongest \rooD  \
deviation, additional information is needed in order to clarify an
underlying orbital effect.

\subsection{Temperature dependence of the crystal structure in
antiferromagnetic \re\ ($RE$ = La, Nd, Sm )}

\begin{figure}[t]
\begin{center}
\includegraphics*[width=1.00\columnwidth]{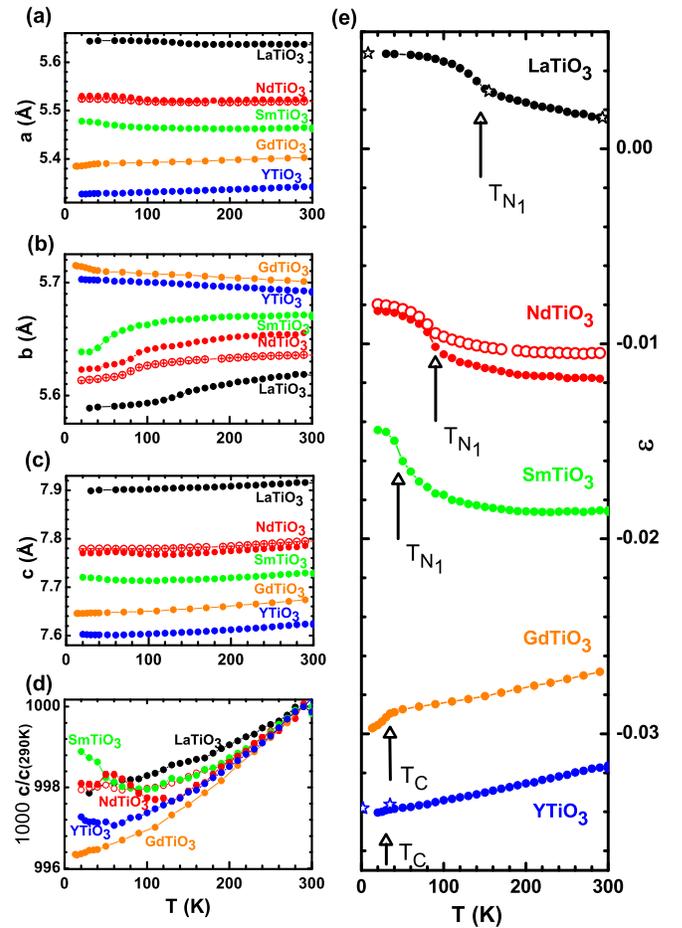}
\end{center}
\caption{(color online) (a-c) Orthorhombic lattice parameters of
$RE$TiO$_3$ ($RE$=La, Nd, Sm, Gd and Y) as a function of
temperature. (d) $c$ parameter scaled to the room temperature
value. (e) Orthorhombic splitting $\varepsilon=(a-b)/(a+b)$. The
arrows indicate the magnetic transition temperatures T$_{N1}$ /
T$_C$ and the stars mark the values for $\varepsilon$ measured by
powder neutron diffraction. Open circles denote the values for the
second Nd$_{1-x}$TiO$_{3+\delta}$ sample with a different
stoichiometry and T$_N$ = 81 K. } \label{fig2}
\end{figure}

For the $RE$ titanates with larger \rei\ showing antiferromagnetic
order, we observe a strong rise of the \emph{a} lattice parameter
at the onset of the magnetic ordering on cooling. Concomitantly,
the \emph{b} lattice parameter decreases more rapidly leading to a
strong increase in the orthorhombic splitting
$\varepsilon=(a-b)/(a+b)$ upon cooling. This anomalous behaviour
of the \emph{a} lattice parameter was first observed for La\tit\
\cite{cwik,hemberger}. Our measurements on other \re\ show that
this anomalous rise of $\varepsilon$ becomes even enhanced for
smaller $RE$ in the series of \re\ as long as the system stays
antiferromagnetic, see Fig. \ref{fig2}(e).

\par

\begin{figure}[t]
\begin{center}
\includegraphics*[width=0.95\columnwidth]{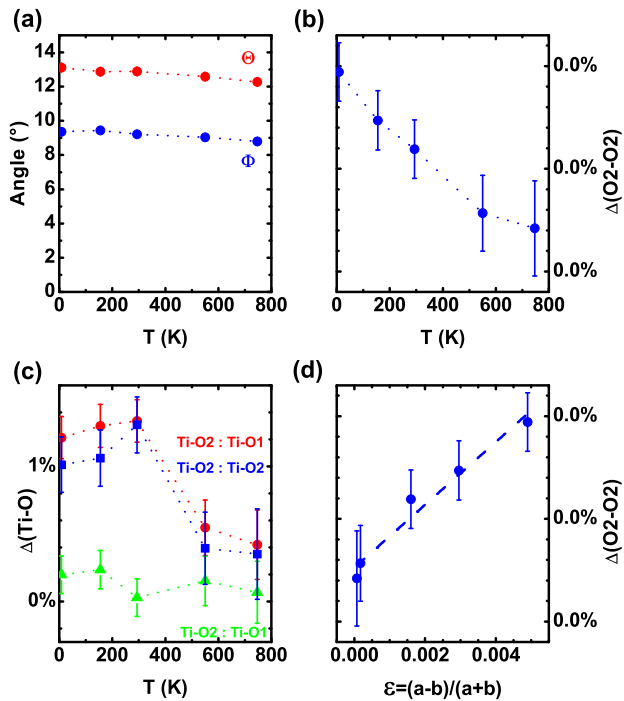}
\end{center}
\caption{(color online) Results of our powder neutron diffraction
measurements on La\tit\ at different temperatures. The first three
points were taken from Ref. \cite{cwik}. The others were measured
on the same sample. (a) \tioct \ octahedral tilt $\Theta$ and
rotation $\Phi$. (b) Differences of the O2-O2 bond lengths. (c)
Differences of the Ti-O bond lengths. (d) Differences of the O2-O2
bond lengths versus $\varepsilon$. The approximately linear
dependence of $\Delta$(O2-O2 bond length) and $\varepsilon$
underlines that $\varepsilon$ is a good indicator for the
octahedral distortion $\Delta$(O2-O2 bond length). (Lines are
guide to the eyes.)} \label{fig3}
\end{figure}

In Fig. \ref{fig3} the results of powder neutron diffraction
measurements on La\tit\ up to T=750\ K are shown. The structure
data at the first three temperatures were taken from Ref.
\cite{cwik} where the same sample as in the present study was
used. Over the full temperature range, the variation of both the
octahedral tilt $\Theta$ and the rotation $\Phi$ is very small,
but the ratio of the O2-O2 octahedron edge lengths changes
essentially, from approximately 4\% at 10\ K to approximately half
of this value at 750\ K. Thus only the internal octahedral
distortions vary strongly with temperature. The decoupling of tilt
and rotation angles on the one side and of the octahedron
distortion on the other side demonstrates that the latter is not
just a consequence of the former. The structural bond-length
mismatch does not drive the octahedral distortion. Instead, the
structural anomaly has to be ascribed to an orbital effect which,
however, does not break any symmetry but just enhances the
crystal-field splitting promoted through the tilt and rotation
distortions.

Since the octahedral tilt $\Theta$ is nearly temperature
independent, the rise of the orthorhombic splitting $\varepsilon$
can be taken as an indicator for the elongation of the \tioct \
basal plane \roo , because only $\Theta$ and this distortion
define the value of $\varepsilon$. Indeed, the plot of \roo \
versus $\varepsilon$ shows a linear dependence (see Fig.
\ref{fig3}).

 As can be seen in Fig. \ref{fig2}, the low-temperature increase
of $\varepsilon$  is more pronounced for the \re\ with smaller
$RE$, Nd\tit \ and Sm\tit, still showing antiferromagnetic order.
Furthermore, with decreasing \rei\ the rise of $\varepsilon$
starts closer to the magnetic ordering temperature, even though it
sets in well above \TN \ for all samples studied. In Sm\tit \ with
a \TN \ of only \TNSm\ the anomalies are most pronounced. The
increase of $\varepsilon$ by 0.01 between 100\ K and 10\ K
indicates an enhancement of the octahedral distortion
characterized  by \roo \ by 1\% , which is quite remarkable for an
internal structure parameter. For Sm\tit \ the temperature
dependence of the crystal structure has been studied by
single-crystal X-ray diffraction down to 100\ K. Again there is
almost no variation in the rotation and tilt angles whereas the
internal octahedral deformation significantly changes. Compared to
La\tit \ the increase of \roo \ between 300 and 100\ K is much
smaller but most likely much stronger between 100 K and lowest
temperatures, see Table \ref{tableX}. Qualitatively, the observed
structural anomalies at the antiferromagnetic ordering also agree
with a mechanism based on spin-orbit coupling \cite{kugel}, since
the elongation of the octahedron occurs parallel to the ordered
spin moment pointing along the $a$ direction, but the observed
effect appears to be too large.

As can be seen in Fig. 3, we have studied a second sample of
Nd$_{1-x}$TiO$_{3+\delta}$, which due to minor inhomogeneity
exhibits a lower N\'eel transition at $T_N$=81\ K. Clearly, the
structural anomalies are strongly suppressed in this material.

In La\tit\ no structural anomaly is seen along the $c$ direction,
see also Ref. \cite{hemberger}, but Nd\tit\ and Sm\tit\ exhibit
such anomalies at the N\'eel ordering. Upon passing into the
antiferromagnetic state the lattice expands along the $c$
direction. The relative strength of these anomalies is, however,
still much lower than those appearing in the $ab$ planes. Since
the ground-state orbital \cite{cwik,pavarini} in La\tit\ has
significant extension along $c$ this elongation agrees still with
the orbital-ordering effect.


\subsection{ Temperature dependence of the crystal structure in
Ferri-/Ferromagnetic \re\ ($RE$ = Gd, Y)}

The temperature dependences of the lattice constants $a$ and $b$
in Gd\tit \ and in Y\tit \ are totally different from that in the
titanates with antiferromagnetic ordering discussed above. The
ferromagnetic compounds show a negative thermal-expansion
coefficient along the $b$ direction and a larger positive one
along $a$. Again, there are well-defined anomalies around the
magnetic ordering, which, however, exhibit the opposite signs
compared to the antiferromagnetic compounds. The anomalies are
visible in Fig. \ref{fig2}. In the case of Y\tit \ these
low-temperature structural anomalies were already reported in
reference \cite{nakao}. The anomalies are, however, much stronger
for Gd\tit.

The crystal structure of Y\tit \ was studied by  neutron
diffraction on a powder sample at 35 K, i.e. slightly above \TC,
and at 2 K and by single-crystal X-ray diffraction on a different
sample between 100 K and room temperature.
Besides the changes in the lattice constants discussed above, no
significant structural changes could be detected between 100\ K
and room temperature. Furthermore, the powder data at 35 and 2\ K
did not reveal a measurable change in the crystal structure as
well. The Ti-O2 bond-length distortion is rather large in Y\tit ,
\rmoD =2.91\% (at 100 K), whereas the other parameter is small,
\roo =1.08\% (at 100 K). Again, the rotation and the tilt angles
vary very little between room temperature and 2 K. The structural
anomalies observed in these ferromagnetic titanates cannot be
explained by the spin-orbit coupling. The most prominent
octahedron elongation occurs along the $b$ direction, whereas one
would expect the spin-orbit coupling to cause an elongation along
the $c$ direction which is the direction of the ferromagnetic
ordered moment \cite{kugel}. Note, however, that indeed there is
such an elongation along the $c$ direction upon passing in the
ferromagnetic state. In addition to the spin-orbit coupling, the
orbital-ordering effect may also contribute to the $c$-axis
expansion, as the lowest orbital level in YTiO$_3$ also possesses
significant extension along $c$
\cite{pavarini,solovyev1,solovyev2}.

\subsection{Magnetostriction and thermal-expansion measurements}

\begin{figure}[t]
\begin{center}
\includegraphics*[width=0.75\columnwidth]{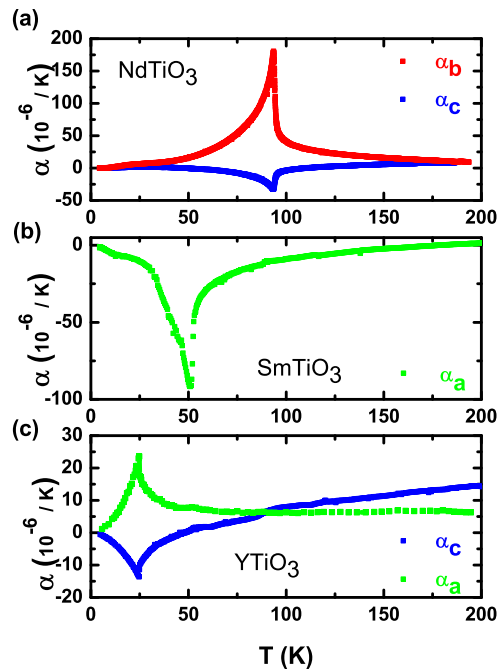}
\end{center}
\caption{(color online) Thermal expansion $\alpha=1/l\cdot\partial
l/\partial T$ for (a) Nd\tit\ parallel to $b$ and $c$, (b) Sm\tit\
parallel to $a$ and (c) Y\tit\ parallel to $a$ and $c$.}
\label{figTherm}
\end{figure}

In order to better resolve the structural anomalies, we have
measured the thermal-expansion coefficients
$\alpha=1/L\cdot\partial L/\partial T$ for three \re\ compounds
with $RE$ = Nd, Sm and Y (see Fig. \ref{figTherm}) on a
high-resolution capacitance dilatometer \cite{lorenz97a}. The
results are given in Fig. \ref{figTherm} and should be compared to
the similar measurement on La\tit\ described in reference
\cite{hemberger}. For these studies we have used almost fully
untwinned single crystals, as it can be ascertained by the
comparison between the dilatometer and the diffraction results.

As expected from the diffraction results, the thermal-expansion
anomalies in Sm\tit \ and in Nd\tit \ are much stronger than those
in La\tit ; in particular the anomaly in the $c$ parameter of
Nd\tit \ is well resolved,  whereas it has no counterpart in the
high-resolution studies on La\tit \ \cite{hemberger}.  The
high-resolution studies in Sm\tit \ and in Nd\tit \ confirm our
conclusion that the structural anomalies set in far above the
N\'eel  ordering \cite{note-aa}.  Furthermore, the smaller
anomalies at the ferromagnetic ordering in Y\tit \ are confirmed
by the dilatometer study. In Y\tit, the $c$-axis anomaly is larger
and the anomaly along the $a$ direction is reversed in comparison
to the antiferromagnetic compounds.

\begin{figure}[t]
\begin{center}
\includegraphics*[width=0.75\columnwidth]{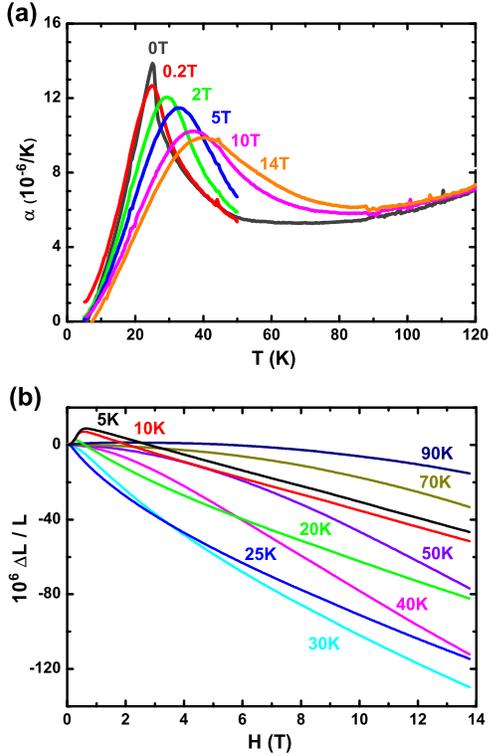}
\end{center}
\caption{(color online) (a) Thermal expansion
$\alpha=1/l\cdot\partial l/\partial T$ of Y\tit\ for various
magnetic fields as a function of temperature ($H \| a$). (b)
Magnetostriction $\Delta L(H)/L$ of Y\tit\ as a function of
applied magnetic field $H \| a$.} \label{figMagneto}
\end{figure}

In Fig. \ref{figMagneto} we show the results of magnetostriction
experiments on YTiO$_3$. The measurement was performed approximately
along the
$a$-direction \cite{note-bb}. The anomaly in the thermal-expansion
coefficient broadens and shifts to higher temperature with
increasing magnetic field. This field dependence is expected for a
ferromagnetic transition. Applying the magnetic field at low
temperature well in the ferromagnetic phase, we first observe an
elongation along the $a$ direction of about ${\Delta a \over a}
\sim 1\cdot$10$^{-5}$. At zero field the magnetic moments in Y\tit
\ order ferromagnetically along the $c$ axis; under an external
field along the $a$ direction they first flip parallel to the
field. The small elongation at small field can hence be associated
with the reorientation of the ordered moment. This elongation is
most likely caused by the spin-orbit coupling elongating the
octahedron along the ordered moment \cite{kugel}. This low-field
magnetostriction is, however, smaller than the magnetostriction
occurring at higher field. The high-field effects have to be
considered as an enhancement of the integrated zero-field
thermal-expansion anomalies, as it can be seen in Fig.
\ref{figMagneto} a). At zero field the ferromagnetic ordering in
Y\tit \ is accompanied by a small antiferromagnetic moment along
the $a$ axis due to the competing magnetic interactions
\cite{ulrich}, which are determined through the orbital
arrangement. Upon cooling in zero field, the orbital arrangement
is changing in the way that ferromagnetism is further stabilized;
this effect seems to be further strengthened by the external field
which anyway stabilizes ferromagnetic order.

\subsection{Crystal structure across the $RE$\tit  \ series}

\begin{figure}[t]
\begin{center}
\includegraphics*[width=0.85\columnwidth]{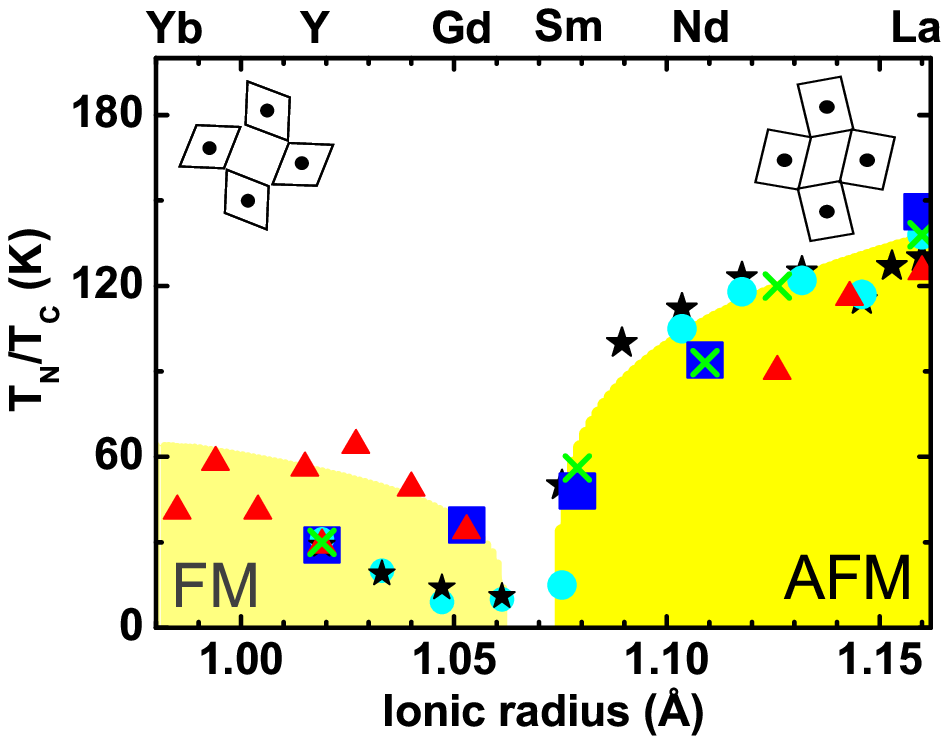}
\end{center}
\caption{(color online) Magnetic phase diagram for \re\ and
La$_{1-x}$Y$_{x}$TiO$_3$. T$_C$ and T$_N$ are plotted as a
function of the \rei. Squares: T$_C$/T$_N$ for \re\ from this
work, triangles: values for \re\ from Greedan \cite{greedan},
crosses: values for \re\ from Katsufuji \emph{et al.}
\cite{katsufuji}, circles: values for La$_{1-x}$Y$_{x}$TiO$_3$
from Okimoto \emph{et al.} \cite{okimoto}, stars: values for
La$_{1-x}$Y$_{x}$TiO$_3$ from Goral \emph{et al.} \cite{goralLaY}.
The inserts show the principal octahedral deformations of the end
members. } \label{figPhase}
\end{figure}

Fig. \ref{figPhase} shows the magnetic phase diagram of $RE$\tit\
as a function of the \rei . In addition to the pure $RE$\tit \
samples we have included results on samples where the A site is
occupied by a mixture of La and Y. The diagram clearly shows how
the crossover from ferromagnetic to antiferromagnetic order is
driven by the \rei , even though some influence of additional
effects is visible. In the ferromagnetic $RE$\tit , there is a
stronger variation of the Curie temperature which indicates the
direct influence of the $RE$-Ti magnetic interaction. Non-magnetic
Y implies a comparably low Curie temperature in this series.
Nevertheless, the \rei \ can be considered as the main external
parameter driving the magnetic transition in the pure as well as
in the mixed compounds.

\begin{figure}[h!]
\begin{center}
\includegraphics*[width=1.00\columnwidth]{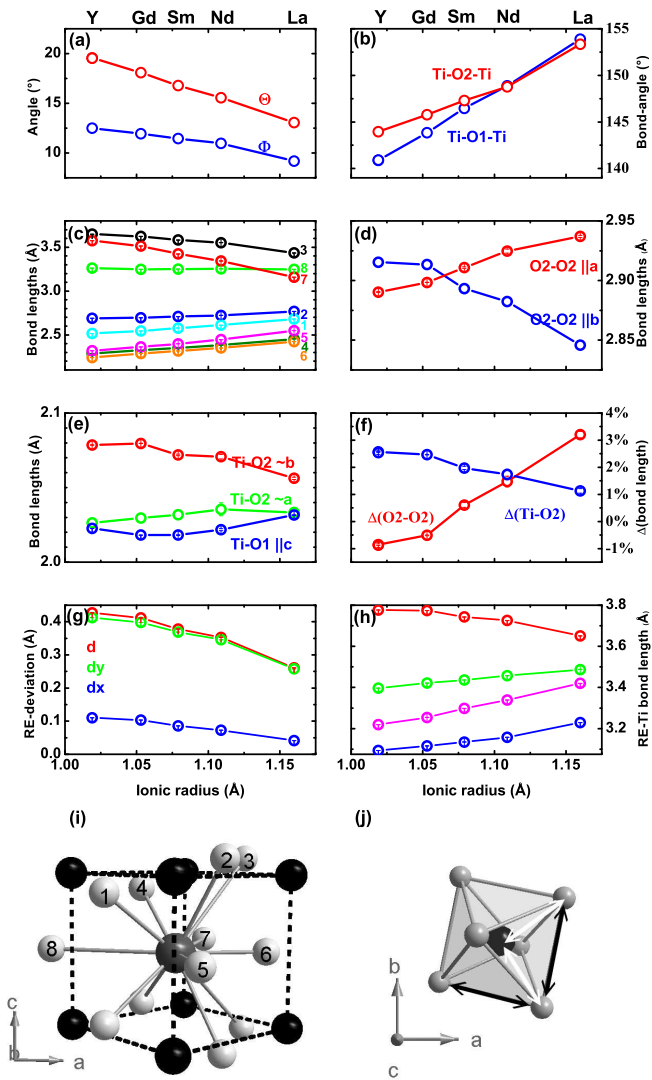}
\end{center}
\caption{(color online) Results from single crystal X-ray
diffraction measurements at room temperature (open circles). (a)
Apical tilt $\Theta$ and basal rotation $\Phi$ of the \tioct \
octahedra of \re. (b) Ti-O-Ti bond angles. (c) $RE$-O bond lengths
(see plot $(g)$ for notation). (d) O2-O2 bond lengths. (e) Ti-O
bond lengths. (f) ratio of O2-O2 bond lengths and ratio of Ti-O2
bond lengths (see also plot $(h)$). (g) Displacement of the $RE$
ion from the symmetry position $0 0 ${\tiny$\frac{1}{4} $}. (h)
$RE$-Ti bond lengths. (i) Oxygen coordination of the $RE$$^{3+}$
ion with the numbering  for plot $(c)$. (j) \tioct\ - black arrows
indicate both O2-O2 bonds which are about parallel to $a$ or $b$
and white arrows indicate the two Ti-O2 bonds which are rotated by
$\Phi$ towards $a$ or $-b$.} \label{fig4}
\end{figure}

In Fig. \ref{fig4} the results of our single crystal diffraction
measurements performed at room temperature are resumed.
Qualitatively, these results agree with previous studies
\cite{maclean} but there are significant quantitative differences
due to the improved sample quality and due to he higher precision
rendered possible with the modern CCD-X-ray techniques. 

Both Nd\tit\ crystals with T$_N$ of 94 K and of 81 K were studied
at room temperature. 
The lower T$_N$ of the second crystal 
already indicates some non-stoichiometry. 
With X-ray diffraction it is difficult to determine 
precisely the oxygen content, therefore, it is not astonishing
that the change in the refined oxygen content per formula unit
from 2.952(30) (high-T$_N$ sample) to 2.988(12) (low-T$_N$ sample)
is not very significant. However, in the low-T$_N$ sample we find 
a significant amount of vaccancies on the Nd-site with a content
per formula unit of only 0.979(1),
whereas the high-T$_N$ sample as
well as the other $RE$TiO$_3$ compounds with high transition temperatures
show the ideal $RE$ to Ti ratio within comparable precision.
The structural parameters obtained for the low-T$_N$ and high-T$_N$ samples
are comparable in view of the large variation of the internal parameters 
in the $RE$TiO$_3$ series, see Table I, however clear differences can be detected. 
The main influence of the excess
oxygens consists in a reduction of the octahedron tilting. Since
excess oxygen and Nd-vacancies enhance the Ti valence and thereby reduce the
effective Ti-ionic radius, this tilt reduction is in perfect
agreement with the bond-length mismatch scenario described by the
tolerance factor. A similar suppression of the tilt distortion by
excess oxygen was also found in
La$_2$CuO$_{4+\delta}$\cite{bradenLSCO}. 
The enhanced Ti-valency in the non-stoichiometric sample is  directly 
observed in the bond-valence sum which is 0.045(1) larger than that in the 
stoichiometric compound.

In Fig. \ref{fig4} we
plot the structural parameters against the rare-earth ionic
radius. The largest overall structural changes concern the tilt
and the rotation deformations. The decrease of the \rei \ enhances
the bond-length mismatch and these angles, as seen in Fig.
\ref{fig4} $(a)$ and in the Ti-O-Ti angles shown in Fig.
\ref{fig4} $(b)$. The distortion of the ideal perovskite structure
gets extremely strong for Y\tit \ which appears close to the
stability of the $Pbnm$ structure. Due to increasing tilt and
rotation deformations the $RE$-O distances vary strongly within
the series; the ratio of the longest to the shortest bond
increases from 1.42 in La\tit \ to 1.63 in Y\tit . For a rigid
tilt an O1 site directly moves towards a $RE$ site yielding the
shortest $RE$-O bond in the entire series ($RE$-O distance \#6 in
Fig. \ref{fig4}). This short bond seems to cause the inversed
orthorhombic splitting seen in many weakly distorted perovskites,
as discussed above. For the smaller rare earths, however, the
systems seems to avoid the extremely short $RE$-O1 bond distance
by shifting the $RE$ site. As can be seen in Fig. \ref{fig4} $(g)$
the shift of the $RE$ site with respect to its high-symmetry
position at (0.5,0,0.25) varies stronger than the oxygen internal
parameters related with the tilting. Due to this $RE$ shift, the
shortest $RE$-O1 distance does not get very small and at the same
time the distance to the opposed O1 sites increases less, see Fig.
\ref{fig4} $(c)$. Furthermore, the $RE$-site shift in $y$
direction results in a strong splitting of the two distances to
the two O1 sites near (0.5,$\pm$0.5,0.25). The structural change
implied through the smaller ionic radius in the \re  \ series is,
therefore, not only described by the increasing tilt and rotation
angles, but in addition there is a change in the $RE$-O
coordination.

With the increasing shift of the $RE$ site for smaller \rei \ the
deformation of the $RE$-Ti coordination increases as well, as it
is shown in Fig. \ref{fig4} $(h)$. In the cubic perovskite, each
$RE$ has eight nearest Ti neighbors; these distances significantly
split for the smaller \re \ most likely destabilizing the orbital
arrangement of the LaTiO$_3$  type \cite{cwik,pavarini}.

For the Ti-orbital physics the internal parameters of the \tioc \
octahedra are important. In agreement with the previous discussion
\cite{cwik} there is a crossover in the octahedral distortion
through the \re  \ series. Whereas La\tit \ exhibits the
rectangular elongation of the basal plane characterized by \roo,
Y\tit \ exhibits the rhombic distortion characterized through
distinct in-plane Ti-O bond distances and \rmoD. As stressed in
reference \cite{cwik}, this structural crossover is accompanied by
the magnetic crossover from the antiferromagnetic to the
ferromagnetic ground state. A part of this crossover seems to be
induced by the structural effects discussed above, i.e. the
response of the $RE$-O arrangement to the extremely strong tilt
and rotation distortions. For the smaller rare earths the
displacement of the $RE$ site is more important and has a stronger
direct influence on the crystal-field splitting of the \tg  \
levels. This purely structural explanation of part of the
deformation is further supported by the similar effects occurring
in the $RE$FeO$_3$ series, whose structure was studied in detail
\cite{refeo1,refeo2,refeo3}. The $RE$FeO$_3$ series shows almost
the identical tilt and rotation angles as a function of rare-earth
ionic radius causing similar constraints in the $RE$-O
coordination. Interestingly also this system exhibits the
octahedral basal-plane elongation for larger rare earths and the
Fe-O2 bond distance splitting for smaller rare earths, but both
effects are much weaker than in the case of the corresponding
titanates.

With the analysis of the intrinsic structural response to the
smaller \rei \ we may more deeply discuss the structural anomalies
occurring in all $RE$\tit \ studied as a function of temperature.
It appears that the tilt, the rotation and the $RE$-site shifts
implied through the \rei \ already favor a certain crystal-field
splitting \cite{pavarini,moshizuki}, which in case of La\tit \
results in a ferroorbital ordering (within the $ab$-plane) whereas
it causes an antiferroorbital ordering in Y\tit . The compounds
close to the structural crossover, where the nature of the lowest
orbital level should change as well, appear most interesting. The
ferroorbital and the antiferroorbital ordering imply an
antiferromagnetic and a ferromagnetic nearest-neighbor magnetic
interaction, respectively. Close to the structural and magnetic
crossover the magnetic interaction is hence significantly reduced
in agreement with the lower values of \TC \ and \TN\  (see Fig.
\ref{figPhase}). Upon cooling, each of the titanates passes into a
well ordered state and in order to suppress the magnetic
fluctuations the magnetic interaction parameters should get
enhanced on the cost of a lattice deformation. Since either the
ferromagnetic or the antiferromagnetic interaction should increase
upon cooling, either the antiferroorbital ($RE$= Gd, Y) or the
ferroorbital ($RE$=La, Nd, Sm) ordering (within the $ab$-plane)
must get enhanced by rotating the orbital degree of freedom. This
explains first the sign change of the in-plane anomalies in the
series and second the observation of the strongest anomalies in
the compounds close to the crossover. Upon continuously varying
the magnetic nearest-neighbor interaction $J$, it has to pass zero
which due to the diverging magnetic fluctuations yields an
energetically unfavorable phase. In all compounds, the orbital
effects follow the crystal-field splitting implied by the
structural deformation, thereby enhancing the magnetic
interaction.

As suggested in the phase diagram shown in Fig. \ref{figPhase} the
\rei \ should be considered as the external parameter driving the
magnetic transition from antiferromagnetic to ferromagnetic
ordering. Therefore, the \re  \ phase diagram may be compared to
the many known systems where a magnetic transition may be implied
by varying some external control parameter. As it has been
recently discussed for a generic quantum phase transition one may
expect a sign change in the thermal-expansion anomalies when
crossing the transition \cite{garst} due to the fact that the
system will avoid the accumulation of entropy just at the
transition. The sign changes in the thermal-expansion anomalies
were for example observed across the metamagnetic transition in
Ca$_{2-x}$Sr$_x$RuO$_4$ \cite{kriener}.
\begin{table*}
\begin{ruledtabular}
\begin{center}
 \centering{ {\scriptsize
\begin{tabular}[t]{c}
 \\
\begin{tabular}[t]{ccccccccccccc}
\textbf{$RE$:} & &  \textbf{\emph{La$\ddag$}}            & \textbf{\emph{Nd$\dag$}} & \textbf{\emph{Nd*$\dag$}} & \textbf{\emph{Sm}}   &  \textbf{\emph{Sm}} & \textbf{\emph{Sm}} & \textbf{\emph{Gd}} & \textbf{\emph{Y}} & \textbf{\emph{Y}}   \\
\textbf{T (K)} & &  \textbf{\emph{290}}         & \textbf{\emph{290}}      & \textbf{\emph{290}}       & \textbf{\emph{290}}  &  \textbf{\emph{130}} & \textbf{\emph{100}} & \textbf{\emph{290}} & \textbf{\emph{290}} & \textbf{\emph{200}}  \\
 \\\emph{Data:}\\
\textbf{obs. refl.}  & & 21502 & 24349                 & 16509                 & 34784           & 4265         & 4830         & 18460        & 23944       & 4779   \\
\textbf{av. refl.}  & & 1557   & 1516                  & 1076                  & 4507            & 1814         & 2202         & 4139         & 2925        & 721     \\
\textbf{redund.}  & & 13.81  & 16.06                   & 15.34                 & 7.72            & 2.35         & 2.19         & 4.46         & 8.19        & 6.63     \\
\textbf{R$_{int}$}  & & 2.68\% & 3.23\%                & 2.45\%                & 2.83\%          & 2.26\%       & 2.13\%       & 2.41\%       & 2.85\%         & 2.84\%   \\
\textbf{2$\Theta_{max}$}  & & 76\grad\  & 107.7\grad\  & 68.8\grad\            & 132.3\grad\     & 84.1\grad\   & 95.1\grad\   & 134.3\grad\  & 112.6\grad\     & 79.1\grad\   \\
 \\\emph{Lattice:}\\
\textbf{a (\AA)} & &  5.63676(25) & 5.52532(87)        & 5.51917(15)           & 5.4647(2)       &  5.4643(2)   & 5.4651(2)    & 5.4031(2)    & 5.3425(2)   & 5.3210(7)    \\
\textbf{b (\AA)} & &  5.61871(26) & 5.65945(88)        & 5.63593(16)           & 5.6712(2)       &  5.6669(2)   & 5.6626(2)    & 5.7009(2)    & 5.6925(2)   & 5.6727(7)   \\
\textbf{c (\AA)} & &  7.91615(34) & 7.79066(121)       & 7.79506(22)           & 7.7291(3)       &  7.7154(3)   & 7.7133(3)    & 7.6739(2)    & 7.6235(2)   & 7.5949(10)    \\
 \\\emph{Atoms:}\\
\textbf{x(RE1)} & &  0.99265(1) &  0.98691(2)          &  0.98788(1)           &  0.98441(1)     &  0.98471(3)  &  0.98473(2)  &  0.98092(1)  &  0.97937(2)  &  0.97873(4)     \\
\textbf{y(RE1)} & &  0.04587(3) &  0.06103(2)          &  0.05788(1)           &  0.06501(1)     &  0.06476(3)  &  0.06488(3)  &  0.06985(1)  &  0.072475(2)  &  0.07324(4)     \\
\textbf{z(RE1)} & &  0.25      &  0.25                 &  0.25                 &  0.25           &  0.25        &  0.25        &  0.25        &  0.25       &  0.25                \\
\textbf{x(Ti1)} & &  0         &  0                    &  0                    &  0              &  0           &  0           &  0           &  0          &  0                    \\
\textbf{y(Ti1)} & &  0.5       &  0.5                  &  0.5                  &  0.5            &  0.5         &  0.5         &  0.5         &  0.5        &  0.5                 \\
\textbf{z(Ti1)} & &  0         &  0                    &  0                    &  0              &  0           &  0           &  0           &  0          &  0                  \\
\textbf{x(O1)} & &  0.0809(2)  & 0.0962(3)             & 0.0932(1)             & 0.1026(1)       &  0.1008(5)   & 0.1010(4)    & 0.1102(2)    & 0.1188(1)   & 0.1195(3)      \\
\textbf{y(O1)} & &  0.49079(18)  & 0.4792(4)           & 0.4787(3)             & 0.4725(2)       &  0.4743(4)   & 0.4742(4)    & 0.4662(2)    & 0.4587(1)   & 0.4587(3)       \\
\textbf{z(O1)} & &  0.25       & 0.25                  & 0.25                  & 0.25            &  0.25        & 0.25         & 0.25         & 0.25        & 0.25              \\
\textbf{x(O2)} & &  0.70975(11)  & 0.7005(2)           & 0.7013(1)             & 0.6975(1)       &  0.6975(3)   & 0.6976(3)    & 0.6943(1)    & 0.6909(1)   & 0.6908(2)         \\
\textbf{y(O2)} & &  0.29348(13)  & 0.3019(2)           & 0.3006(1)             & 0.3039(1)       &  0.3035(3)   & 0.3037(3)    & 0.3064(1)    & 0.3085(1)   & 0.3090(2)        \\
\textbf{z(O2)} & &  0.04210(9)  & 0.0487(2)            & 0.0472(1)             & 0.0515(1)       &  0.0514(2)   & 0.0516(2)    & 0.0541(1)    & 0.0576(1)   & 0.0577(2)       \\
\emph{U$_{ij}$ (\AA $^2$):}\\
\textbf{U$_{11}$(RE1)} & &  0.00463(6) & 0.00740(4)    & 0.00633(4)            & 0.00500(2)      &  0.0053(1)   & 0.00497(6)   & 0.00732(2)   & 0.00517(4)    & 0.0041(1)    \\
\textbf{U$_{22}$(RE1)} & &  0.00470(6) & 0.00671(5)    & 0.00608(5)            & 0.00427(2)      &  0.0038(1)   & 0.00364(6)   & 0.00626(2)   & 0.00391(4)    & 0.0036(1)   \\
\textbf{U$_{33}$(RE1)} & &  0.00456(5) & 0.00674(5)    & 0.00528(5)            & 0.00445(2)      &  0.0048(1)   & 0.00452(7)   & 0.00731(2)   & 0.00523(4)    & 0.0041(1)   \\
\textbf{U$_{12}$(RE1)} & &  -0.00077(2) & -0.00069(3)  & -0.00087(3)           & -0.00071(1)     &  -0.00059(4) & -0.00055(3)  & -0.00046(1)  & -0.00041(3)   & -0.0003(1)   \\
\textbf{U$_{13}$(RE1)} & &  0 &           0            & 0                     & 0               &  0           & 0            & 0            & 0             & 0             \\
\textbf{U$_{23}$(RE1)} & &  0 &           0            & 0                     & 0               &  0           & 0            & 0            & 0             & 0              \\
\textbf{U$_{11}$(Ti1)} & &  0.00226(18) & 0.00541(9)   & 0.0034(1)             & 0.00308(4)      &  0.0035(2)   & 0.0036(1)    & 0.00576(6)   & 0.00375(7)   & 0.0027(2)   \\
\textbf{U$_{22}$(Ti1)} & &  0.00258(19) & 0.00517(11)  & 0.0021(1)             & 0.00311(4)      &  0.0030(2)   & 0.0029(2)    & 0.00561(6)   & 0.00398(7)   & 0.0033(2)   \\
\textbf{U$_{33}$(Ti1)} & &  0.00249(15) & 0.00500(12)  & 0.0031(1)             & 0.00241(4)      &  0.0032(2)   & 0.0030(2)    & 0.00553(6)   & 0.00325(7)   & 0.0019(2)   \\
\textbf{U$_{12}$(Ti1)} & &  -0.00005(6) & -0.00006(9)  & -0.00005(9)           & 0.00000(3)      &  -0.00002(13) & -0.00007(10) & -0.00005(4)  & -0.00014(5)   & -0.0002(1)   \\
\textbf{U$_{13}$(Ti1)} & &  -0.00005(5) & 0.00001(7)   & -0.00012(6)           & -0.00005(3)     &  -0.00004(12) & -0.00008(10)  & -0.00004(4)  & -0.00025(5)   & -0.0003(1)   \\
\textbf{U$_{23}$(Ti1)} & &  -0.00020(8) & -0.00002(10) & -0.00033(9)           & 0.00002(3)      &  -0.00001(15) & -0.00004(11) & 0.00012(4)   & 0.00006(5)   & -0.0000(1)   \\
\textbf{U$_{11}$(O1)}  & &  0.0073(5) & 0.00951(46)    & 0.0089(4)             & 0.0070(2)       &  0.0081(9)   & 0.0080(7)    & 0.0092(3)    & 0.0084(3)   & 0.0052(7)    \\
\textbf{U$_{22}$(O1)}  & &  0.0097(5) & 0.00904(58)    & 0.0104(6)             & 0.0072(2)       &  0.0051(9)   & 0.0043(7)    & 0.0087(3)    & 0.0070(3)   & 0.0050(7)    \\
\textbf{U$_{33}$(O1)}  & &  0.0055(4) & 0.00500(59)    & 0.0061(6)             & 0.0034(2)       &  0.0049(9)   & 0.0046(7)    & 0.0064(3)    & 0.0059(3)   & 0.0022(8)    \\
\textbf{U$_{12}$(O1)}  & &  -0.0002(3) & -0.00128(45)  & 0.0026(3)             & -0.0013(2)      &  -0.0017(7)  & -0.0014(5)   & -0.0013(2)   & -0.0015(2)   & -0.0010(5)    \\
\textbf{U$_{13}$(O1)}  & &  0 &          0             & 0                     & 0               &  0           &   0          &   0          &   0           &   0          \\
\textbf{U$_{23}$(O1)}  & &  0 &          0             & 0                     & 0               &  0           &   0          &   0          &   0           &   0           \\
\textbf{U$_{11}$(O2)}  & &  0.0064(3) & 0.0082(3)    & 0.0069(2)             & 0.0056(1)         &  0.0065(6)   & 0.0057(5)    & 0.0082(2)    & 0.0076(2)   & 0.0045(5)   \\
\textbf{U$_{22}$(O2)}  & &  0.0072(3) & 0.0075(4)    & 0.0037(4)             & 0.0053(2)         &  0.0041(6)   & 0.0041(5)    & 0.0076(2)    & 0.0069(2)   & 0.0042(5)   \\
\textbf{U$_{33}$(O2)}  & &  0.0087(3) & 0.0090(5)    & 0.0090(4)             & 0.0070(2)         &  0.0066(6)   & 0.0069(6)    & 0.0095(2)    & 0.0091(3)   & 0.0055(6)   \\
\textbf{U$_{12}$(O2)}  & &  -0.0018(2) & -0.0018(3)  & -0.0012(2)            & -0.0016(1)        &  -0.0015(5)  & -0.0009(4)   & -0.0014(2)   & -0.0016(2)   & -0.0002(1)   \\
\textbf{U$_{13}$(O2)}  & &  0.0008(2) & 0.0007(3)    & 0.0014(2)             & 0.0009(1)         &  0.0005(4)   & 0.0007(3)    & 0.0009(2)    & 0.0014(2)   & 0.0009(4)   \\
\textbf{U$_{23}$(O2)}  & &  -0.0074(2) & -0.0009(4)  & -0.0016(3)            & -0.0014(1)        &  -0.0011(5)  & -0.0010(4)   & -0.0014(2)   & -0.0017(2)   & -0.0011(4)   \\
 \\\emph{Fit:}\\
\textbf{GoF}  & &           2.29      & 2.40          & 1.22                  & 2.09            & 2.54         & 2.56         & 2.33         & 2.11        & 2.07       \\
\textbf{R}  & &              1.32\%    & 2.11\%        & 1.62\%                & 1.34\%          & 1.95\%       & 1.99\%       & 1.82\%       & 1.70\%      & 1.89\%    \\
\textbf{R$_w$}  & &          2.97\%    & 3.87\%        & 4.30\%                & 2.46\%          & 4.77\%       & 4.96\%       & 3.53\%       & 2.53\%      & 3.92\%   \\
\end{tabular} \\
\\
\end{tabular}
 \\
}} \caption{\label{tableX} Results of single crystal X-ray
diffraction measurements of \re\ ($RE$ = La, Nd, Sm, Gd, Y) at
room temperature and for $RE$ = Sm, Y down to 100 K. *: results
for a (non-stoichiometric) Nd\tit-sample with T$_N$ = 81 K.
$\ddag$: The La\tit \ single crystal was twinned with the
following ratios of the six possible domains: $82.7\%, 8.1\%,
4.2\%, 1.8\%, 1.5\%$ and $1.5\%$. $\dag$: Both Nd\tit-crystals
were slightly twinned ($a\leftrightarrow b$) with a major twin
domain of roughly 90\% and all reflections with $|h-k|\leq 1$
($h,k\neq0$) were excluded as they could not be separated from
other twin contributions in the integration procedure.}
\end{center}
\end{ruledtabular}
\end{table*}
\subsection{La$_{1-x}$Y$_x$TiO$_3$ (x=0.50)}

\begin{figure}[h!]
\begin{center}
\includegraphics*[width=0.95\columnwidth]{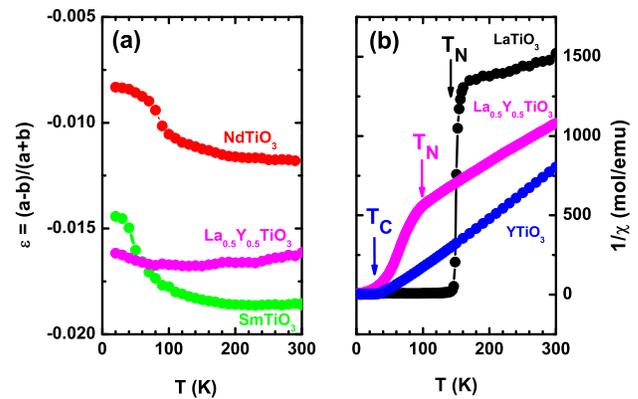}
\end{center}
\caption{(color online) (a) Orthorhombic splitting $\varepsilon$
of La$_{1-x}$Y$_x$\tit\ (x=0.50) compared with the other \re\ with
similar rare earth ionic radii. (b) Inverse magnetic
susceptibility $1/\chi$ of La$_{0.5}$Y$_{0.5}$\tit\ compared with
La\tit, Y\tit\ and the average of La\tit\ and Y\tit. Arrows
indicate T$_C$/T$_N$.} \label{fig6}
\end{figure}

Since the \rei \ drives the structural deformations, the orbital
arrangement and the magnetic structure, it would be quite
interesting to study this phase diagram by continuously varying
the ionic radius. La$_{1-x}$Y$_x$TiO$_3$ was thought to be a good
system since there is no magnetic moment on the $RE$ site. Several
groups have already studied this system, see references
\cite{goralLaY,katsufuji}. Indeed this system shows the transition
from antiferromagnetism to ferromagnetism between 30 and 40 \% of
Y \cite{goralLaY}. We have prepared a powder sample of
La$_{0.50}$Y$_{0.50}$TiO$_3$ and show the inverse of its magnetic
susceptibility in Fig. \ref{fig6} $(b)$. There is no well-defined
anomaly comparable to the effects in the pure $RE$\tit\ but the
weak effect around 90 K indicates the onset of antiferromagnetic
ordering in qualitativ agreement with the Curie-Weiss analysis of
the susceptibility at higher temperatures. From these magnetic
properties and from the averaged ionic radius which is between
those of Nd and Sm one would expect similar physical properties as
those observed in Nd\tit\ and in Sm\tit. Fig. \ref{fig6} $(a)$
shows, however, that this is not the case. Although the
orthorhombic lattice parameters are in between the corresponding
values of the pure Nd and Sm compounds, comparable structural
anomalies at low temperatures do not occur in
La$_{0.50}$Y$_{0.50}$TiO$_3$, see the plot of the orthorhombic
distortion in Fig. \ref{fig6} $(a)$. We think that the disorder
arising from the mixed occupation of La and Y with quite different
ionic radii destroys the effects induced by orbital ordering.
Although this result annihilates the hope to tune a mixed titanate
to the critical averaged \rei\ and to drive it thereby into a
critical configuration, it corroborates our conclusion that the
structural anomalies in the $RE$\tit\ materials arise from orbital
physics.

\section{Conclusion.}

The detailed structural analysis of the $RE$\tit \ series as a
function of temperature allows us to separate the structural
effects implied directly through the variation of the \rei \ from
those related with an orbital effect. The decrease of the \rei \
leads first to the enhancement of the tilt and rotation angles and
second to a modification of the $RE$-O and $RE$-Ti coordinations.
For small distortion angles the $RE$-site is closer to its
high-symmetry position whereas it moves away from that for the
larger distortions. As has been stated by several groups these
shifts yield already a significant crystal field splitting of the
\tg  \ levels. The character of the thereby lowered orbital is
rather different and the orbital arrangement, which is either
ferroorbital or antiferroorbital in nature, is even opposed for
the end-members of the series.

The temperature dependence of the crystal structure, however,
shows that anomalous structural deformations occur in all $RE$\tit
\ which should modify the character of the orbital with the lowest
energy. Since the ferroorbital and antiferroorbital $ab$-plane
arrangement is related with an antiferromagnetic and ferromagnetic
nearest-neighbor interaction, respectively, the structural change
directly couples to the magnetism.  The structural anomalies are
clearly associated with the magnetic ordering temperatures where
the thermal-expansion coefficients exhibit extrema. We, therefore,
conclude that magnetism drives a change in the orbital ground
state. In all compounds, this change points in the sense to
enhances the dominant magnetic interaction. For antiferromagnetic
$RE$\tit \ with La, Nd and Sm the ferroorbital arrangement is
strengthened while for Gd\tit \ and Y\tit \ the antiferroorbital
arrangement is strengthened. In this sense the temperature-driven
orbital ordering follows the crystal-field splitting already
imposed through the strong distortions. The structural effects are
weakest in the end members and strongest close to the structural
and magnetic crossover. Unfortunately, it seems not possible to
drive the titanates into the critical configuration by
continuously varying the ionic radius in a mixed system
La$_{1-x}$Y$_x$TiO$_3$ as disorder effects seem to suppress
orbital degrees of freedom.
 \\
 \\
\section{Acknowledgemets.}
This work was supported by the Deutsche Forschungsgemeinschaft
through Sonderforschungsbereich 608. We thank D. Khomskii for
valuable discussions, N. Schittner and N. Hollmann for various
susceptibility measurements and D. Meier, J. Rohrkamp and O. Heyer
for some thermal expansion experiments.

\begin{table}[!h]
\centering{ {\scriptsize
\begin{ruledtabular}
\begin{tabular}[t]{cccccc}
\textbf{RE:} & &  \textbf{\emph{La}} & \textbf{\emph{La}} & \textbf{\emph{Y}} & \textbf{\emph{Y}} \\
\textbf{T (K)} & &  \textbf{\emph{550}} & \textbf{\emph{747}} & \textbf{\emph{2}} & \textbf{\emph{35}} \\
 \\
 \textbf{M$_c$ ($\mathbf{\mu_B/Ti^{3+}}$)} &  & 0 & 0  & 0.753(75) & 0 \\
  \\
 \emph{lattice:}\\
\textbf{a (\AA)} &  & 5.6386(2) & 5.6454(2) & 5.3226(1) & 5.3239(1)  \\
\textbf{b (\AA)} & &  5.6367(2) & 5.6447(2) & 5.6952(1) & 5.6944(1)  \\
\textbf{c (\AA)} & &  7.9518(2) & 7.9699(3) & 7.5962(2) & 7.5952(2)  \\
 \\\emph{atoms:}\\
\textbf{x(RE1)} & &  0.9928(5) &  0.99329(1)  &  0.9776(2)  &  0.9780(2)    \\
\textbf{y(RE1)} & &  0.0387(2) &  0.0353(3))  &  0.0740(1)  &  0.0738(2)    \\
\textbf{z(RE1)} & &  0.25      &  0.25       &  0.25       &  0.25             \\
\textbf{x(Ti1)} & &  0         &  0          &  0          &  0               \\
\textbf{y(Ti1)} & &  0.5       &  0.5        &  0.5        &  0.5              \\
\textbf{z(Ti1)} & &  0         &  0          &  0          &  0               \\
\textbf{x(O1)} & &  0.0778(5)  & 0.0759(7)   & 0.1213(2)   & 0.1212(2)      \\
\textbf{y(O1)} & &  0.4882(4)  & 0.4885(5)   & 0.4570(2)   & 0.4572(2)      \\
\textbf{z(O1)} & &  0.25       & 0.25        & 0.25        & 0.25           \\
\textbf{x(O2)} & &  0.7102(4)  & 0.7113(5)   & 0.6901(1)   & 0.6901(2)      \\
\textbf{y(O2)} & &  0.2908(4)  & 0.2896(5)   & 0.3092(2)   & 0.3095(2)      \\
\textbf{z(O2)} & &  0.0401(2)  & 0.0391(3)   & 0.0577(1)   & 0.0576(1)      \\
\emph{B (\AA $^2$):}\\
\textbf{B(RE1)} & &  1.07(2) & 1.33(3) & 0.177(12) & 0.193(12)  \\
\textbf{B(Ti1)} & &  0.55(3) & 0.70(4) & 0.190(25) & 0.181(26)  \\
\textbf{B(O1)}  & &  1.05(3) & 1.32(4) & 0.257(16) & 0.305(17)  \\
\textbf{B(O2)}  & &  1.21(3) & 1.60(3) & 0.228(10) & 0.264(10)  \\
 \\\emph{fit:}\\
\textbf{R}  & & 4.31\% & 4.45\% & 3.61\% & 3.70\%  \\
\textbf{R$_w$}  & & 5.78\% & 5.93\% & 4.89\% & 4.91\%  \\
\end{tabular} \\
\end{ruledtabular}
}} \caption{\label{tableN} Results of the rietveld refinement of
powder neutron diffraction measurements of \re\ ($RE$ = La and Y)
at different temperatures.}
\end{table}


\begin{thebibliography}{}
\bibitem{TbAlO3}We excluded TbAlO$_3$ as the
atomic positions were determined by powder X-ray diffraction with
the consequence of larger errors.
\bibitem{note-aa} The overall agreement between the macroscopic
thermal expansion experiment and the diffraction results indicates
that these samples are nearly untwinned, with the minority domains
containing about  10\% of the total volume.
\bibitem{note-bb} Most likely due to an imperfect orientation,
about 15\% of other directions are superposed explaining the
weaker anomaly at T$_C$ when compared to the thermal expansion
experiment.
\end{thebibliography}
\end{document}